\documentclass[aps,pra,floatfix,noshowkeys,epsfig,graphics]{revtex4}
\usepackage{graphicx}
\usepackage{amsmath}
\usepackage{amsfonts}
\usepackage{amssymb}
\usepackage{comment}
\usepackage{color}

\definecolor{revisiongreen}{rgb}{0,0.45,0}

\newcommand{\newc}{\newcommand}

\newc{\beq}    {\begin{equation}}
\newc{\eeq}    {\end{equation}}
\newc{\EV}{~{\rm eV}}

\newc{\beqa}    {\begin{eqnarray}}
\newc{\eeqa}    {\end{eqnarray}}
\newc{\bs}    {\section}
\newc{\no}    {\\ \nonumber}

\def\apj{{\em Astrophys. J.  }}

\def\mnras{{ Mon. Not. Roy. Astron. Soc.  }}

\newcommand{\bea}{\begin{eqnarray}}
\newcommand{\eea}{\end{eqnarray}}

\newc{\st}    {\stackrel}

\begin{document}
\title{Alleviating   the Hubble tension with self-interacting  ultralight dark matter}

\author{Jae-Weon Lee}
\email{scikid@jwu.ac.kr}
\affiliation{Department of Electrical and Electronic Engineering, Jungwon University, 85 Munmu-ro, Goesan-eup, Goesan-gun, Chungcheongbuk-do, 28024, Korea.}

\begin{abstract}
We study the possibility that a self-interacting ultralight dark matter field with a characteristic energy scale $\tilde{m}\simeq 1~{\rm eV}$ can behave like an extra radiation component before recombination. This additional radiation can reduce the sound horizon of the photon–baryon fluid. A smaller sound horizon can give a higher inferred value of the present-day Hubble parameter at the background level, and thus may help reduce the Hubble tension. This effect can be obtained without adding another dark sector component besides self-interacting ULDM. A full perturbation and likelihood analysis is needed to see if the model is consistent with CMB and structure-formation data.
\end{abstract}

\maketitle

\section{Introduction}

The $\Lambda$CDM model, which contains  cold dark matter (CDM) and a cosmological constant $\Lambda$, has been very successful in explaining many observations of the universe. However, some tensions with recent observational data still remain~\cite{Efstathiou:2024dvn}.
The Hubble tension~\cite{Verde:2019ivm,Kamionkowski:2022pkx,DiValentino:2021izs,Knox:2019rjx,10.1093/mnras/stad3357}, one of the biggest mysteries in modern cosmology, refers to the discrepancy between direct measurements of the current Hubble parameter $H_0$, such as those based on supernova observations, and values inferred from the $\Lambda$CDM model and cosmic microwave background (CMB) data.

Measurements of the angular separation of acoustic peaks in the CMB  
\beq  
\label{theta}
\theta_s = \frac{r_s}{D_M}
\eeq
 provide strong constraints on $H_0$, as the parameter $ \theta_s = (1.04112 \pm 0.00029) \times 10^{-2} $
  is tightly constrained in the Planck 2015 TT,TE,EE+lowP+lensing+ext base-$\Lambda$CDM fit~\cite{Planck:2015fie}. We retain its central value only as a fixed benchmark for the background comparison.
Here, $r_s$ refers to the comoving sound horizon radius
of the photon-baryon fluid, whereas
\beq
D_M = \frac{c}{H_0} \int_0^{z_{\text{ls}}} \frac{dz}{\left[\rho(z)/\rho_0\right]^{1/2}},
\eeq
denotes 
the transverse comoving distance in a spatially flat universe
 to the last scattering surface at redshift $z_{ls}$, where $\rho(z)$ is the total energy density at redshift $z$ and $\rho_0$ is its present-day value.
$H_0$ shows a discrepancy between values obtained from local  measurements grouping near  $H_0 \simeq 73 \, {\rm km} \, \text{s}^{-1} \, {\rm Mpc}^{-1}$ and those inferred from CMB observations
grouping near
$H_0\simeq 68 \, {\rm km} \, {\rm s}^{-1} \, {\rm Mpc}^{-1}$
in the $\Lambda$CDM model, resulting in a difference of 5\% to 9\% depending on the data sets used
\cite{Wang:2023bxf,Verde:2023lmm,Scolnic:2024hbh}.

There are two  main categories of proposed solutions. In early-time solutions, a decrease in $r_s$ by introducing   early dark energy~\cite{Poulin:2018cxd}  or extra radiation~\cite{Vagnozzi:2019ezj}
requires a decrease in $D_M$ at fixed $\theta_s$. This can favor a larger $H_0$, but the inferred shift depends on the other cosmological parameters and the late-time expansion history.
Such early components must subsequently become sufficiently subdominant to preserve the successful late-time evolution. Some proposed solutions also exacerbate the $S_8$ tension
~\cite{Pedrotti:2024kpn,Hill:2020osr,Vagnozzi:2023nrq}.
On the other hand, late-time solutions leave $r_s$ essentially unchanged and modify the late-time expansion history, for example through phantom dark energy or new interactions in the dark sector, so that the required $D_M$ can be maintained with a larger $H_0$
\cite{Montani:2024pou}. However, such modifications are strongly constrained by BAO and supernova data~\cite{Cai:2021weh,Krishnan:2021dyb,Keeley:2022ojz,Jiang:2024xnu,Benevento:2020fev}.

There are many other tensions encountered by the $\Lambda$CDM model
at galactic scales.
These include small-scale issues~\cite{2000PhRvL..85.1158H,Salucci:2002nc,1996ApJ...462..563N,2003MNRAS.340..657D,2003IJMPD..12.1157T}, the typical size and mass of galaxies~\cite{Lee:2008ux}, the satellite plane problem~\cite{Park:2022lel}, and the final parsec problem~\cite{koo2024final,Bromley:2023yfi}, among others. Ultralight dark matter (ULDM), also known as fuzzy dark matter, scalar field dark matter,  or the ultralight axion
with mass $m\simeq 10^{-22}~{\rm eV}$
~\cite{1983PhLB..122..221B,Sin:1992bg,Lee:1995af,Matos:1998vk,2000PhRvL..85.1158H,Boehmer:2007um}, is a compelling alternative to CDM.
The extremely low mass permits a large phase-space occupation, and ULDM can be described as a coherent classical field that, for suitable production mechanisms and initial conditions, may
alleviate
the problems of the $\Lambda$CDM model.
While on large scales, its perturbations mimic CDM, on smaller scales, quantum pressure suppresses structure formation.
It preserves the success of CDM on scales larger than galaxies while potentially resolving its challenges at galactic scales (for a review, see ~\cite{Hui:2016ltb,Ferreira_2021,Rindler_Daller_2021, Lee_2018,Matos_2024}).

However, ULDM with mass $m\simeq 10^{-22}~{\rm eV}$ and without
self-interaction (i.e., fuzzy dark matter) itself faces some tensions with observations, such 
as those from
the Lyman-alpha forest~\cite{Irsic:2017yje,Armengaud:2017nkf}. Introducing self-interaction of ULDM 
~\cite{Lee:1995af,Boehmer:2007um,Chavanis:2011zi}
can modify the small-scale phenomenology~\cite{Dave:2023wjq,2024arXiv241210285L}. However, repulsive self-interaction also produces pressure that suppresses structure formation, and its strength is constrained by cosmological and halo observables~\cite{Hartman:2021upg,Shapiro:2021hjp}.
Self-interacting ULDM has also been proposed as a mechanism to address 
the dynamical-friction timing problem of globular clusters in the Fornax dwarf galaxy~\cite{Koo:2025jkx},
and 
to generate neutrino masses and potentially relate them to the electroweak scale~\cite{Lee:2024rdc}.

In this paper, we propose that self-interacting ULDM 
with a quartic interaction
possesses unique features that may make it a suitable  candidate for an early-time solution to
the Hubble tension.
If the field oscillates in an effective quartic potential  before the matter-radiation equality, it can behave as a dark radiation-like component, reducing the sound horizon and helping to alleviate the Hubble tension.
There are proposals relying on additional ultralight fields~\cite{Ye:2021iwa,Flambaum:2019cih,Gonzalez:2020fdy,Agrawal:2019lmo}; however, these models often require an extra field that is not the primary dark matter component, while
in our model ULDM
is the main dark matter.

In our scenario, the reduction of the sound horizon can lead to a higher inferred value of $H_0$. Since the ULDM component enters its matter-like phase before recombination, its impact on the CMB may be reduced compared with models containing a substantial radiation-like component at recombination. However, the detailed CMB acoustic structure must ultimately be tested using a full perturbation and Boltzmann analysis not discussed in this work.
This paper is a proof-of-concept, order-of-magnitude study of the homogeneous background mechanism, not a precision cosmological fit. We use a simple piecewise evolution to illustrate the characteristic scales and the possible direction and size of the effect. 

In Sec.~II, we briefly investigate the cosmological evolution of self-interacting ULDM. In Sec.~III, we examine how its temporary radiation-like phase can reduce the sound horizon and alleviate the Hubble tension. We summarize and discuss the implications of our results in Sec.~IV.

\section{self-interacting Ultralight dark matter}

To understand the evolution of dark matter, 
we consider a ULDM scalar field $\phi$
with an action
\beq
\label{action}
 S=\int \sqrt{-g} d^4x[\frac{-R}{16\pi G}
-\frac{g^{\mu\nu}} {2} \phi^*_{;\mu}\phi_{;\nu}
 -V(\phi)],
\eeq
where the potential for the field 
is
\beq
\label{V}
V(\phi)=\frac{m^2}{2}|\phi|^2 + \frac{\lambda |\phi|^4}{4}
\eeq 
with  a dimensionless self-coupling constant $\lambda>0$~\cite{Lee:1995af,Chavanis:2011zi,Boehmer:2007um}.
This model is a minimal, renormalizable extension of the free ULDM model.
A typical energy scale of $\phi$,
\beq
\tilde{m}\equiv\frac{m}{ \lambda^{1/4}}
\eeq
is crucial for our analysis~\cite{Li:2013nal,Boudon:2022dxi}.
For a repulsively self-interacting condensate in the Thomas--Fermi (TF) limit,
the static ground state density profile is proportional to
$\sin(kr)/(kr)$~\cite{Boehmer:2007um,Lee:1995af,Diez_Tejedor_2014}. Its first zero defines the Thomas--Fermi radius
$R_{\rm TF}=\pi/k$. For the potential
$V(\phi)=m^2|\phi|^2/2+\lambda|\phi|^4/4$ and the definition
$\tilde m\equiv m/\lambda^{1/4}$, this radius becomes~\cite{Diez_Tejedor_2014}
\begin{equation}
\label{RTF}
R_{\rm TF}
=\frac{\pi m_{\rm P}}{\tilde m^2}
=48.92~{\rm kpc}
\left(\frac{1~{\rm eV}}{\tilde m}\right)^2,
\end{equation}
where $m_{\rm P}=2.435\times10^{27}~{\rm eV}$
is the reduced Planck mass.
For nine representative galaxies in the 2019 SPARC analysis~\cite{2019RoAJ...29..109C}, the fitted static halo radii $R_{\rm TF} \sim  4.281$--$42.376~\rm kpc$ correspond to $\tilde{m} = 1.07$--$3.38~\rm eV$ under the Thomas–Fermi relation. This is a range of individual fit values, not a statistical confidence interval or a universal observational bound.
The stellar kinematics of eight classical Milky Way dwarf spheroidal
galaxies favor 
$R_{\rm TF}\sim0.5$--$2~{\rm kpc}$, corresponding to
$\tilde m\sim4.95$--$9.89~{\rm eV}$ under the normalization adopted
here~\cite{Diez_Tejedor_2014}. 
However, the applicability of these estimates is limited because the dark matter halos of real dwarf galaxies need not follow the ground state profile, and the inferred halo parameters can depend sensitively on the model-dependent treatment of visible matter.
For comparison, Eq.~(\ref{RTF}) refers to the stationary complex-field Thomas--Fermi halo solution. Its direct application to halos formed from the zero-charge oscillating cosmological branch requires additional dynamical assumptions.
Nevertheless, these galactic analyses motivate considering {\rm eV}-scale values of $\tilde{m}$.

The evolution of the field is described by the Klein-Gordon equation 
\beq
\square \phi-2 \frac{d V}{d|\phi|^2} \phi=0,
\eeq
where $\square$ is the d'Alembertian.
For the homogeneous background we take a constant complex phase and vanishing conserved $U(1)$ charge, so that $\phi$ can be chosen real. In this sector the equation becomes
\beq
\ddot{\phi} + 3H\dot{\phi} + m^2\phi + \lambda \phi^3 = 0,
\label{eom}
\eeq
which describes the cosmological evolution of the coherent background ULDM field. Here, the
dot represents a derivative with respect to time, and $H$ 
is the Hubble parameter at
that time.
The persistence and stability of the coherent zero-charge background against self-resonance and transverse fluctuations are assumed here and require a separate analysis.

For a rapidly and coherently oscillating scalar field in a potential
$
V(\phi)\propto |\phi|^n,
$
the time-averaged virial relation gives
$
\langle K\rangle=\frac{n}{2}\langle V\rangle
$, where $K$ denotes kinetic energy density of the field.
The averaged equation-of-state parameter is therefore~\cite{PhysRevD.28.1243}
\beq
w=\frac{\langle K\rangle-\langle V\rangle}{\langle K\rangle+\langle V\rangle}
=\frac{n-2}{n+2} \,.
\eeq

The averaged energy density $\rho^\phi$ of the field evolves with the equation-of-state parameter $w$ as $\rho^\phi\propto a^{-3(1+w)}$.
For a quartic potential, \(n=4\), and hence
$w=\frac{1}{3},$
which implies
$
\rho^\phi\propto a^{-4}.
$
Thus, the coherently oscillating ULDM field behaves as a radiation-like component while the quartic term dominates. When the quadratic term becomes dominant, \(n=2\), giving
$w=0$
and
$
\rho^\phi\propto a^{-3},
$
corresponding to pressureless matter.

\begin{figure}[ht]
\includegraphics[width=0.5\textwidth]{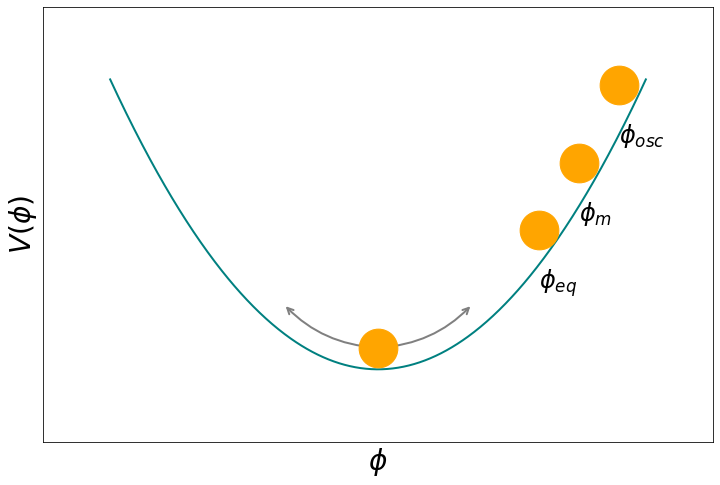}
\caption{ A schematic of the potential $ V(\phi) $ of the ULDM field oscillating in the potential is shown. $ (\phi_{osc}, \phi_m, \phi_{eq})$ here represent the positive oscillation amplitudes (turning-point values) at the redshifts $ (z_{osc}, z_m, z_{eq}) $, respectively.
The oscillating field behaves
as a radiation-like component for $z_m<z<z_{osc}.$}
\label{phi}
\end{figure}

\begin{figure}[ht]
\includegraphics[width=0.8\textwidth]{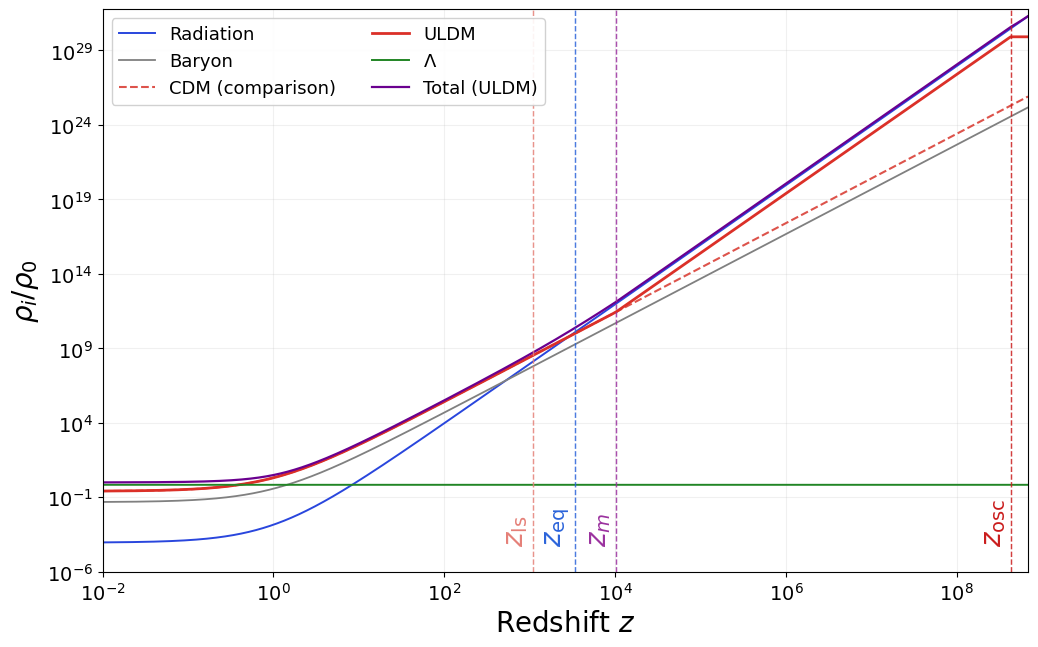}
\caption{Piecewise background densities normalized to the present total density $\rho_{\rm tot,0}$ at $\tilde m=1.5~{\rm eV}$ and $T_{osc}=0.1~{\rm MeV}$. The solid red curve is ULDM, the dashed red curve is CDM with the same present abundance, the blue and gray curves are radiation and baryons, the green horizontal line is the cosmological constant, and the purple curve is the ULDM total. We fix $\omega_b=0.022$, $\omega_\phi=0.119$, $\omega_r=4.2\times10^{-5}$ and $\omega_\Lambda=0.3212$ in this illustration. Vertical lines indicate $z_{ls}$, $z_{eq}$, $z_m$ and $z_{osc}$. Constant post-annihilation $g_*$ and $g_{*s}$ are assumed in drawing the curves; their continuation around and above the onset is schematic and does not represent a BBN computation.}

\label{Tcfig}
\end{figure}

We can describe the evolution of the
background field in three phases. (See Fig. 1.)  
In Phase I ($z \ge z_{osc}$), the field is nearly frozen due to Hubble friction, and $\phi$ behaves as early dark energy ($w \simeq -1$), whose fractional energy density is negligible sufficiently early, although it grows as radiation redshifts.
As the universe expands, the field begins to roll down and oscillate.  
In Phase II ($z_m \le z < z_{osc}$), the quartic term in $V(\phi)$ dominates over the quadratic mass term, and the energy density of the oscillating field decreases like radiation, i.e., $w \simeq 1/3$ and $\rho^\phi \sim (1 + z)^{4}$.  
During this phase, the ULDM field contributes more to the energy density than CDM in the  $\Lambda$CDM model and reduces the sound horizon. 
Fuzzy dark matter ($\lambda=0$), on the other hand, is not applicable to our model because it cannot fulfill the role of radiation.
In Phase III ($z < z_m$), the mass term becomes dominant over the quartic term, and the oscillating field behaves like CDM ($w = 0$). 
This describes the asymptotic background behavior; it does not imply that the perturbations immediately become identical to those of CDM.
Fig. 1 shows a schematic illustration of the potential $V(\phi)$ and the representative oscillation amplitudes.

We approximate the crossover by a sudden transition at equality of the quadratic and quartic contributions evaluated at the oscillation amplitude. The actual averaged pressure changes continuously and remains nonzero near this point. This approximation can introduce a non-negligible systematic error in the sound-horizon reduction and the conditional $H_0$ shift. For simplicity, we do not model the smooth crossover or quantify this error in the present proof-of-concept study. Thus, $T_m$ and $z_m$ below are matching scales in a piecewise background approximation, and the resulting parameter estimates are not precision predictions or observational bounds.

To relate the temperature to the redshift, we use conservation of comoving entropy,
\beq
g_{*s}(T)T^3a^3={\rm const.},
\eeq
which gives
\beq
1+z(T)=\frac{T}{T_0}\left[\frac{g_{*s}(T)}{g_{*s}(T_0)}\right]^{1/3},
\eeq
where $g_{*s}$ is the effective number of entropy degrees of freedom  and $a$ is the 
scale factor.
We examine the broader illustrative range $T_{osc}\lesssim0.1~{\rm MeV}$ and require $T_{osc}\gg T_m$ for a prolonged quartic oscillatory phase. For the post-annihilation examples we approximate $g_*\simeq3.38$ and $g_{*s}\simeq3.93$ between $T_{osc}$ and $T_{eq}$; within this approximation $1+z=T/T_0$. The upper temperature then corresponds to $1+z_{osc}\lesssim 4.26\times10^8$. 
Close to the upper onset limit, the residual temperature dependence of these functions should be retained for a precision BBN calculation which is beyond the scope of this work.
To derive the redshifts below we use
$1+z(T)\simeq T/T_0$,
where $T_0=2.35\times10^{-4}~{\rm eV}$ is the photon temperature of the universe today.

In Phase II, the equation of motion can be approximated
as $\ddot{\phi}+\lambda \phi^3=0$ and the oscillation
time scale is ${\cal O}(1/(\sqrt{\lambda}\phi))$.
The field begins to oscillate when its characteristic frequency becomes comparable to the total Hubble expansion rate. In the quartic-dominated regime we retain the onset convention $H(T_{osc})\simeq\sqrt{\lambda}\phi_{osc}$. Writing 
\beq
r_{osc}\equiv\rho^\phi_{osc}(T_{osc})/\rho_{\rm rad}(T_{osc}),
\eeq
 we use $H(T_{osc})=H_{\rm rad}(T_{osc})\sqrt{1+r_{osc}}$, where
\beq
H_{\rm rad}(T)=\left(\frac{\pi^2g_*(T)}{90}\right)^{1/2}\frac{T^2}{m_P}
\eeq
is the Hubble parameter due to radiation density at $T$.
We neglect the other matter components at the late onset. The relation $\rho^\phi_{osc}\simeq\lambda\phi_{osc}^4/4$ applies to an approximately stationary field while the quartic term dominates. Together with the piecewise density evolution below, it gives $r_{osc}\simeq f_\phi T_{eq}/T_m$ when $g_*$ and $g_{*s}$ are constant between $T_{osc}$ and $T_{eq}$. Hence,
\begin{equation}
\label{Tosc}
T_{osc}\simeq
\left(\frac{90}{\pi^2g_*(T_{osc})}\right)^{1/4}
\lambda^{1/4}\sqrt{m_P\phi_{osc}}\,(1+r_{osc})^{-1/4}.
\end{equation}
Both the onset convention and the sudden matching between the quartic and quadratic regimes introduce order-unity uncertainties in Eqs.~(\ref{Tosc})--(\ref{Tosc2}).  We use the analytic estimates to describe characteristic scales; the sound-horizon calculation below normalizes the piecewise density to the present dark matter abundance. Near $T_{osc}=0.1~{\rm MeV}$ the constant-$g_{*s}$ assumption used to relate the onset to the late abundance is also an approximation.

Using Eq.~(\ref{Tosc}), the energy density of the field at matter-radiation equality is
\begin{equation}
\rho^\phi_{eq}
=\rho^\phi_{osc}\left(\frac{T_m}{T_{osc}}\right)^4
\left(\frac{T_{eq}}{T_m}\right)^3
=\frac{\rho^\phi_{osc}T_mT_{eq}^3}{T_{osc}^4}
=\frac{\pi^2g_*(T_{osc})}{360}(1+r_{osc})
\frac{\phi_{osc}^2T_mT_{eq}^3}{m_P^2}.
\end{equation}
Here $T_m$ is the temperature at which the quartic term is equal to the quadratic term (i.e., $z=z_m$). This gives
\begin{equation}
\label{phiosc1}
\phi_{osc}
=\left(\frac{360}{\pi^2g_*(T_{osc})(1+r_{osc})}\right)^{1/2}
\sqrt{\frac{\rho^\phi_{eq}m_P^2}{T_mT_{eq}^3}}.
\end{equation}
At $z_m$, Phase III starts, and 
$m^2\phi^2/2\simeq \lambda \phi^4$/4, which gives $\phi=\phi_m
\simeq \frac{\sqrt{2}m}{\sqrt{\lambda}}$.
The field behaves as CDM in this phase, and
 the energy density at $z_m$ is
\beq
\rho^\phi_{m}
\simeq 2 m^2 \phi_m^2/2 =2\frac{m^4}{\lambda}=2\tilde{m}^4=\rho^\phi_{eq} \left( \frac{T_m}{T_{eq}}\right)^3, 
\eeq
where we have used
the relation
$\rho^\phi\propto T^3$ in Phase III.
Thus, $\phi_m\simeq \sqrt{2}\frac{\tilde{m}^2}{m}$, and 
\beq
\label{Tm}
T_m=\left(\frac{2\tilde{m}^4}{\rho^\phi_{eq}}\right)^{1/3} T_{eq}.
\eeq
Within this matching prescription, placing the matter-like branch before equality requires $T_m>T_{eq}\simeq0.79~{\rm eV}$. This is not by itself a sufficient condition for negligible pressure or CDM-like perturbations.
Using $f_\phi\simeq0.844$ and $g_*(T_{eq})\simeq3.38$ below, this condition becomes $\tilde m\gtrsim0.65~{\rm eV}$. For $\tilde m\simeq 1~{\rm eV}$ one obtains $T_m\simeq1.39~{\rm eV}$, so the transition remains close to matter-radiation equality.

At matter-radiation equality, the total matter density rather than the ULDM density alone equals the radiation density,
\beq
\rho^\phi_{eq}+\rho^b_{eq}
=\rho_{\rm rad}(T_{eq}),
\eeq
where $\rho^b$ is the baryon density.
Therefore,
\beq
\label{rhoeq2}
\rho^\phi_{eq}
=f_\phi\,\rho_{\rm rad}(T_{eq})
=f_\phi\frac{\pi^2}{30}g_*T_{eq}^4,
\qquad
f_\phi\equiv\frac{\Omega_\phi}{\Omega_\phi+\Omega_b}.
\eeq
For the fiducial values adopted below, $f_\phi\simeq0.844$.

We use $g_*(T_{eq})\simeq3.38$ and $g_{*s}(T_{eq})\simeq3.93$. With $f_\phi\simeq0.844$ and $T_{eq}=0.79~{\rm eV}$, Eq.~(\ref{rhoeq2}) gives $\rho^\phi_{eq}\simeq0.366~{\rm eV}^4$.
Then, from Eqs.~(\ref{phiosc1}), (\ref{Tm}), and (\ref{rhoeq2}) we obtain
\begin{equation}
\label{phiosc2}
\phi_{osc}
=\frac{2^{5/6}}{\sqrt{1+r_{osc}}}\sqrt{3}\,
f_\phi^{2/3}
\left(\frac{\pi^2g_*(T_{eq})}{30}\right)^{1/6}
m_P
\left(\frac{T_{eq}}{\tilde m}\right)^{2/3},
\end{equation}
where $g_*(T_{osc})\simeq g_*(T_{eq})\simeq3.38$ has been used as an approximation across the expanded onset range. In the adopted matching scheme,
\beq
r_{osc}(\tilde m)\simeq0.479
\left(\frac{\tilde m}{1~{\rm eV}}\right)^{-4/3},
\qquad
\phi_{osc}\simeq\frac{5.84\times10^{18}~{\rm GeV}}{\sqrt{1+r_{osc}(\tilde m)}}
\left(\frac{\tilde m}{1~{\rm eV}}\right)^{-2/3}.
\eeq
Thus, $\phi_{osc}\simeq4.80\times10^{18}~{\rm GeV}$ for $\tilde m=1~{\rm eV}$ and $\simeq1.24\times10^{18}~{\rm GeV}$ for $\tilde m=10~{\rm eV}$. 
This condition determines $\phi_m$ and $\phi_{osc}$ for given $m$ and $\lambda$, although the ultraviolet completion of the present phenomenological potential is not specified here.

Inserting $\rho^\phi_{eq}$ into Eq.~(\ref{Tm}) gives
$T_m\simeq1.39~{\rm eV}\,(\tilde m/1~{\rm eV})^{4/3}$. Since $g_{*s}(T_{m})=g_{*s}(T_0)\simeq3.93$ in this regime, the entropy factor in the temperature-redshift relation is unity, and
\beq
1+z_m\simeq5.92\times10^3
\left(\frac{\tilde m}{1~{\rm eV}}\right)^{4/3}.
\eeq
Using Eqs.~(\ref{Tosc}), (\ref{Tm}), and (\ref{phiosc2}), one also obtains
\begin{eqnarray}
\label{Tosc2}
T_{osc}
=\sqrt{\frac{6}{1+r_{osc}}}\,2^{-1/12}
f_\phi^{1/3}
\left(\frac{\pi^2g_*(T_{eq})}{30}\right)^{-1/6}
m_P\lambda^{1/4}
\left(\frac{T_{eq}}{\tilde m}\right)^{1/3} \\
\simeq 0.0608~{\rm MeV}\,C(\tilde m)
\left(\frac{\lambda}{10^{-91}}\right)^{1/4}
\left(\frac{\tilde m}{1~{\rm eV}}\right)^{-1/3},
\end{eqnarray}
where
$C(\tilde m)\equiv\left[\frac{2}{1+r_{osc}(\tilde m)}\right]^{1/2}$.
At $\tilde m=1~{\rm eV}$, $C\simeq1.163$ and the reference coefficient is $T_{osc}\simeq0.0707~{\rm MeV}$ for $\lambda=10^{-91}$. The illustrative onset condition $T_{osc}\lesssim0.1~{\rm MeV}$ therefore gives
\beq
\lambda\lesssim\frac{7.34\times10^{-91}}{C(\tilde m)^4}
\left(\frac{\tilde m}{1~{\rm eV}}\right)^{4/3},
\qquad
m\lesssim\frac{2.93\times10^{-23}~{\rm eV}}{C(\tilde m)}
\left(\frac{\tilde m}{1~{\rm eV}}\right)^{4/3}.
\eeq
For $\tilde m=1~{\rm eV}$ these become $\lambda\lesssim4.01\times10^{-91}$ and $m\lesssim2.52\times10^{-23}~{\rm eV}$. 
These are approximate onset-scale limits, not observational bounds or evidence that the range passes BBN.
For the stationary complex-field halo comparison, the conventional TF parameter is $\Lambda_{\rm TF}=\lambda M_{\rm Pl}^2/(4\pi m^2)=2\lambda m_P^2/m^2$, where $M_{\rm Pl}=\sqrt{8\pi}m_P$ is the unreduced Planck mass~\cite{Lee:1995af}. The condition $\Lambda_{\rm TF}\gg1$ is a separate halo-regime criterion, not a requirement of the homogeneous background calculation. If imposed together with $T_{osc}\lesssim0.1~{\rm MeV}$, it gives
\begin{equation}
7.11\times10^{-111}
\left(\frac{\widetilde m}{1~{\rm eV}}\right)^4
\ll\lambda\lesssim
\frac{7.34\times10^{-91}}{C(\tilde m)^4}
\left(\frac{\widetilde m}{1~{\rm eV}}\right)^{4/3}.
\end{equation}
A quartic oscillatory interval additionally requires $T_{osc}>T_m$, giving nominally
\begin{equation}
\lambda\gtrsim\frac{2.76\times10^{-110}}{C(\tilde m)^4}
\left(\frac{\widetilde m}{1~{\rm eV}}\right)^{20/3}.
\end{equation}

 Establishing BBN compatibility throughout the enlarged onset range requires a nuclear reaction-network calculation using the actual $\rho^\phi(T)$, the associated expansion rate, and the temperature-dependent standard-model thermodynamics. Such a calculation lies beyond the scope of this paper. 
For $z_{osc}\gg z_m$, the contribution of the onset region to the sound-horizon integral is small. In both numerical figures we fix $T_{osc}=0.1~{\rm MeV}$.

Maintaining both the ultralight scalar mass and its extremely weak self-interaction against quantum and ultraviolet corrections is itself an interesting theoretical question. At a low renormalization scale, the self-interaction corrections to the renormalized potential are perturbatively small. This statement does not protect the ultralight mass against independent ultraviolet threshold contributions.
For an order-of-magnitude estimate, we retain only the radial fluctuation mode as a representative contribution to the one-loop effective potential. The transverse mode of the complex scalar contributes at the same loop order and modifies only numerical coefficients of order unity~\cite{Martin:2013gka}.
The radiative corrections to the scalar potential can be described by the one-loop Coleman--Weinberg effective potential. For the tree-level potential
$
V(\phi)=\frac{1}{2}m^2\phi^2+\frac{\lambda}{4}\phi^4,
$
the field-dependent mass squared is
$
M^2(\phi)=V''(\phi)=m^2+3\lambda\phi^2,
$
and the one-loop Coleman-Weinberg correction is
\beq
V_1(\phi)=\frac{M^4(\phi)}{64\pi^2}
\left[
\ln\left(\frac{M^2(\phi)}{\mu^2}\right)-\frac{3}{2}
\right].
\eeq
For the regime
$3\lambda\phi^2<m^2,$
we define
$
x\equiv\frac{3\lambda\phi^2}{m^2}< 1.
$
Dropping the field-independent term $V_1(0)$ and expanding $\ln(1+x)$,
one obtains
\beq
V_1(\phi)
=
\frac{3\lambda }{32\pi^2}
\left[
\ln\left(\frac{m^2}{\mu^2}\right)-1
\right]m^2\phi^2
+
\frac{9\lambda^2}{64\pi^2}
\ln\left(\frac{m^2}{\mu^2}\right)\phi^4
+
\frac{9\lambda^3}{64\pi^2m^2}\phi^6
+\mathcal{O}(\phi^8),
\eeq
where the
radiative correction to the
quadratic term is a relative $\mathcal{O}(\lambda)$ correction at a low renormalization scale, provided that the logarithm is controlled.
The field-independent contribution $V_1(0)$ does not affect the scalar
equation of motion and is absorbed into the renormalized vacuum energy.
 The first genuinely higher-order interaction generated by the one-loop correction is therefore
$
V_6^{(1)}
=
\frac{9\lambda^3}{64\pi^2m^2}\phi^6.
$
Comparing it with the tree-level quartic interaction
$
V_4=\frac{\lambda}{4}\phi^4,
$
we find
\beq
\frac{V_6^{(1)}}{V_4}
=
\frac{9\lambda^2\phi^2}{16\pi^2m^2}
=
\frac{3\lambda}{16\pi^2}
\left(
\frac{3\lambda\phi^2}{m^2}
\right),
\eeq
which is also negligible. In the opposite regime, 
$3\lambda\phi^2\gg m^2$, 
$M^2(\phi)\simeq 3\lambda\phi^2$,
and the full Coleman--Weinberg correction satisfies
\beq
\frac{V_1(\phi)}{V_4}\sim\frac{9\lambda}{16\pi^2}
\eeq
up to logarithmic factors, which can be controlled by renormalization-group improvement. Hence, for the extremely small quartic coupling relevant to ULDM, radiative corrections remain perturbatively negligible in both regimes \cite{Arvanitaki:2014faa,Sher:1988mj,Machacek:1984zw,Aghaie:2026bqq}
and the self-interaction loop expansion is perturbative. A cutoff estimate instead gives $\delta m^2\sim\lambda\Lambda_{\rm UV}^2/(16\pi^2)$, up to a convention-dependent numerical factor. For a sufficiently high cutoff, including the Planck scale for part of the parameter range, this can exceed $m^2$. A symmetry or other UV protection is therefore an additional assumption, not a consequence of the small low-energy Coleman--Weinberg correction.

\section{Alleviating the Hubble tension}

In this section we investigate the possible background mechanism
based on self-interacting ULDM
for alleviating the Hubble tension.
The comoving sound horizon radius 
can be represented by an integral:
\beq
r_s 
=
\int_{0}^{t_{ls}} \frac{c_s(t)}{a(t)} dt 
= \int_{z_{ls}}^\infty \frac{c_s(z)}{H(z)} dz
\eeq
where $t_{ls}$ is the recombination time, 
\beq
c_s(z) = \frac{c}{\sqrt{3 \left( 1 + \frac{3 \Omega_b}{4 \Omega_\gamma (1 + z)} \right)}}
\equiv \frac{c}{\sqrt{3 \left( 1 +R(z) \right)}}
\eeq
is the sound speed of the photon-baryon fluid.

During Phase II, the scalar density exceeds that of CDM with the same present abundance, raising the pre-recombination expansion rate. The intrinsic crossover to the quadratic regime then makes its background evolution approach the matter-like limit.

We now use the following 
formulas~\cite{Kamionkowski:2022pkx}:
\beq
\label{rseq}
r_s  = \frac{c}{\sqrt{3} H_{\text{ls}}} \int_{z_{\text{ls}}}^\infty \frac{dz}{\left[\rho(z)/\rho(z_{\text{ls}})\right]^{1/2} (1 + R(z))^{1/2}}, 
\eeq
where 
\beq
H_{ls} = 100 \, {\rm km} \, {\rm s}^{-1} \, {\rm Mpc}^{-1} \, \omega_r^{1/2} (1 + z_{ls})^2 \sqrt{1 + \frac{\omega_m}{(1 + z_{ls})\omega_r}} 
\eeq
approximates the expansion rate at $z_{ls}$ in the matter-like branch, neglecting the cosmological constant there. We use $\omega_i=\Omega_i h^2$ as usual, and $\omega_m=\omega_b+\omega_\phi$.
Then,
\beq
\label{H0}
H_0 = \sqrt{3} H_{\text{ls}} \theta_s \frac{ \displaystyle \int_{0}^{z_{\text{ls}}} \left( \dfrac{\rho(z)}{\rho_0} \right)^{-1/2} dz }{ \displaystyle \int_{z_{\text{ls}}}^\infty \left( \dfrac{\rho(z)}{\rho(z_{\text{ls}})} \right)^{-1/2} \dfrac{dz}{\sqrt{1 + R(z)}} }.
\eeq
It is useful to calculate the inverse sound-horizon ratio directly:
\beq
\label{Hratio0}
R_s\equiv\frac{r_s^{\rm (CDM)}}{r_s^{\rm (ULDM)}}=
\frac{\displaystyle\int_{z_{ls}}^\infty [\rho^{\rm (CDM)}(z)]^{-1/2}\dfrac{dz}{\sqrt{1+R(z)}}}
{\displaystyle\int_{z_{ls}}^\infty [\rho^{\rm (ULDM)}(z)]^{-1/2}\dfrac{dz}{\sqrt{1+R(z)}}}.
\eeq
Here $\rho^{\rm (CDM)}$ and $\rho^{\rm (ULDM)}$ denote the total densities in the two cosmologies. Equation~(\ref{Hratio0}) is a ratio of sound horizons, not a ratio of inferred Hubble constants. At fixed $\theta_s$, a reduction in $r_s$ requires the same fractional reduction in $D_M$. However, at fixed $\omega_m$, $D_M$ is not simply proportional to $H_0^{-1}$, so $R_s$ cannot be identified with $H_0^{\rm (ULDM)}/H_0^{\rm (CDM)}$. Determining that ratio requires solving Eq.~(\ref{H0}) with specified parameter constraints, and a comparison with observations additionally requires perturbations and a likelihood analysis.

Fig. 2 shows relative densities of components
as continuous functions of $z$,
especially 
at $z_m$ and $z_{eq}$,
with the total density of the ULDM cosmology given by
\beq
\label{Euldm}
\left[\rho^{\rm (ULDM)}(z)/\rho_{100}\right]^{1/2}= 
\begin{cases}
\sqrt{\omega_r (1+z)^4+\omega_m (1+z)^3+\omega_\Lambda} ~~\text{for $z<z_m$} ~\text{(Phase III)}\\
\sqrt{\omega_r (1+z)^4+\omega_\phi \frac{(1+z)^4}{1+z_m} +\omega_b (1+z)^3+\omega_\Lambda} ~~\text{for $z_m\le z<z_{osc}$} ~\text{(Phase II)} \\
\sqrt{\omega_r (1+z)^4+\omega_b (1+z)^3+\omega_\phi \frac{(1+z_{osc})^4}{1+z_m} +\omega_\Lambda} ~~\text{for $z_{osc}\le z$} ~\text{(Phase I)}, \\
\end{cases} 
\eeq
where $\rho_{100}\equiv\rho_0/h^2$ is the common critical-density normalization. For Fig.~2, we adopt the fiducial values $\omega_b=0.022$, $\omega_\phi=0.119$, $\omega_r=4.2\times10^{-5}$ and $\omega_\Lambda=0.3212$. 
The matter densities are guided by rounded central values from the Planck 2015 TT,TE,EE+lowP+lensing base-$\Lambda$CDM fit ~\cite{Planck:2015fie}, while $\omega_\Lambda$ is chosen to match the reference value of $\theta_s$.

\begin{figure}[ht]
\includegraphics[width=0.70\textwidth]{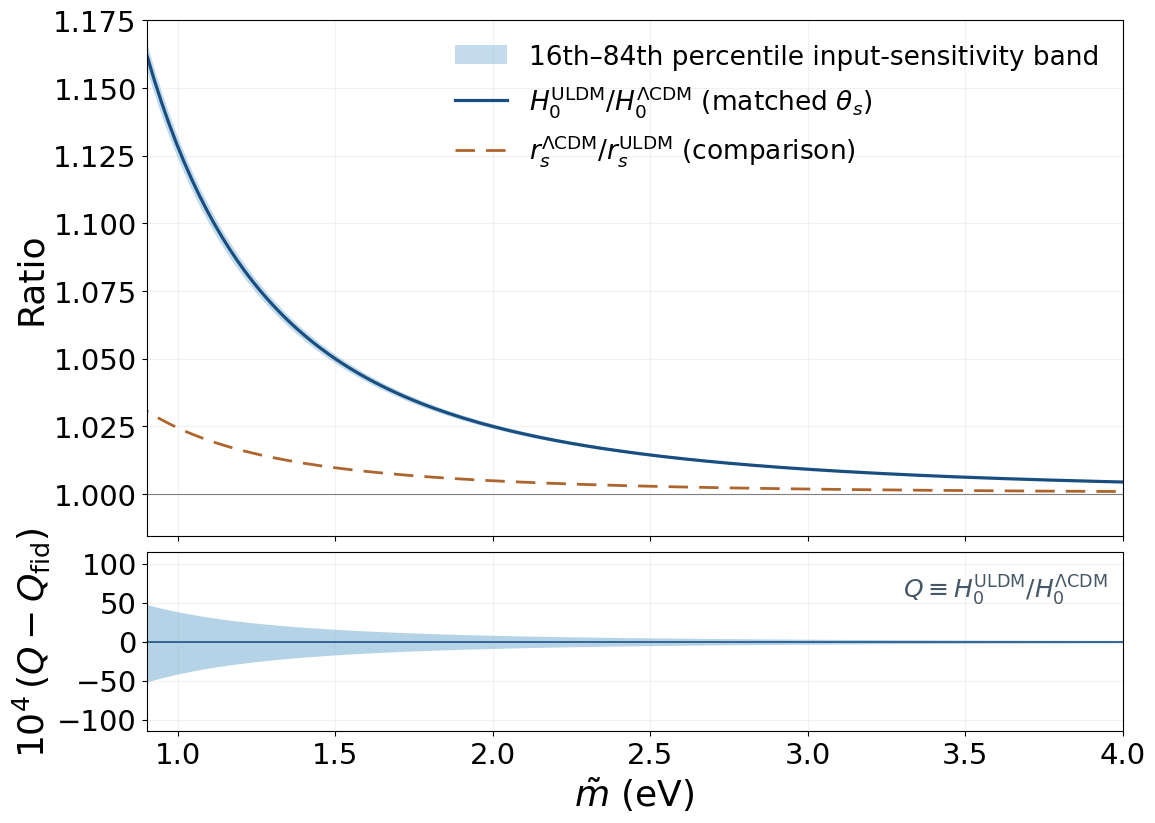}
\caption{The present-day Hubble-parameter ratio
$Q\equiv H_0^{\rm ULDM}/H_0^{\Lambda{\rm CDM}}$ as a function of $\tilde m$
for $T_{\rm osc}=0.1~{\rm MeV}$ (solid blue curve).
The physical densities $\omega_b=0.022$, $\omega_\phi=0.119$ and
$\omega_r=4.2\times10^{-5}$ are fixed at their fiducial values, while
$\omega_\Lambda$ is determined at each mass by matching
$\theta_s=0.0104112$.
The dashed orange curve shows
$R_s=r_s^{\Lambda{\rm CDM}}/r_s^{\rm ULDM}$ for comparison.
The shaded region spans the 16th--84th percentiles from 5000 independent
draws of $\omega_b$ and $\omega_\phi$.
The lower panel shows the deviation of this band from the fiducial
Hubble-parameter ratio $Q_{\rm fid}$ on an enlarged scale.
The band represents sensitivity to the input densities rather than an
observational confidence interval.
The calculation uses a background-only sudden-transition approximation
with $z_{\rm ls}=1089$.}
\label{Hratio}
\end{figure}

Using Eq.~(\ref{Hratio0}), we obtain $R_s\simeq1.03066$ for $\tilde m=0.9~{\rm eV}$ at $T_{osc}=0.1~{\rm MeV}$, corresponding to a $2.97\%$ reduction in $r_s$ in this background approximation. Solving the fixed-$\theta_s$ distance equation gives $Q\equiv H_0^{\rm ULDM}/H_0^{\Lambda{\rm CDM}}\simeq1.16249$ for the same fiducial physical densities.
 For $\tilde m=1.5~{\rm eV}$
 we obtain $R_s\simeq1.00962$
 and $H_0^{\rm ULDM}/H_0^{\Lambda{\rm CDM}}\simeq1.04986$, within the adopted sudden-transition approximation. These are conditional background responses to adjusting $\omega_\Lambda$, not observationally inferred Hubble constants. 
For Fig.~\ref{Hratio}, we set $z_{ls}=1089$, $\omega_\gamma=2.472\times10^{-5}$, $T_0=2.35\times10^{-4}~{\rm eV}$, and $T_{osc}=0.1~{\rm MeV}$, with the other fiducial physical densities given above. 
The value $\omega_\Lambda = 0.3212$ used in Fig.~2 corresponds to the reference $\Lambda$CDM background obtained by matching $\theta_s = 0.0104112$. In Fig.~3, $\omega_\Lambda$ is solved separately for the ULDM background at each mass while keeping the same $\theta_s$.
 At each parameter draw, we use $1+z_{eq}=(\omega_b+\omega_\phi)/\omega_r$, $T_{eq}=T_0(1+z_{eq})$, and $f_\phi=\omega_\phi/(\omega_b+\omega_\phi)$, and recompute $T_m$ from Eq.~(\ref{Tm}). Thus, the derived equality quantities are not varied independently.
For simplicity, neutrinos are treated as effectively massless in both backgrounds, and $z_{ls}=1089$ is held fixed as an illustrative last-scattering redshift. The reference $\Lambda$CDM background therefore does not exactly reproduce the full Planck base-$\Lambda$CDM model.

The input-sensitivity band uses 5000 independent Gaussian draws centered on $\omega_b=0.022$ and $\omega_\phi=0.119$, with standard deviations $0.00023$ and $0.0020$, respectively. These widths are the marginalized $68\%$ uncertainties of the Planck 2015 TT+lowP+lensing base-$\Lambda$CDM fit~\cite{Planck:2015fie}. The centers retain the rounded illustrative values used here. All other inputs are held fixed. The fixed $\theta_s$ benchmark comes from a different data combination specified in the Introduction; these inputs are not a joint fit. Neglecting the parameter covariance makes this a sensitivity illustration, not a sample from the Planck posterior or a confidence interval for this ULDM model. The shaded region contains the pointwise 16th--84th percentiles.

The present calculation concerns only the homogeneous background evolution. During the radiation-like phase, perturbations of the ULDM field need not behave identically to those of standard free-streaming radiation, and they can affect the evolution of gravitational potentials and the CMB acoustic structure. A quantitative test of the scenario therefore requires implementation of the self-interacting ULDM background and perturbation equations in a Boltzmann code such as CLASS or CAMB.

\section{Discussion}
We have explored a scenario in which ULDM behaves as dark radiation
before matter-radiation equality, thereby possibly alleviating the Hubble tension.
The early-time reduction of the sound horizon provides the initial shift, while the numerical $H_0$ response shown here also depends substantially on the late-time adjustment of $\omega_\Lambda$ required to keep $\theta_s$ fixed.
The toy example in this work with $\tilde{m}\simeq 1.5~{\rm eV}$ gives about a $5\%$ increase in $H_0$.
However, our estimate should be regarded only as an order-of-magnitude illustration.
 A nuclear reaction-network calculation with the actual scalar density history is required to test BBN compatibility at this onset temperature and is beyond the scope of this paper. The structure-formation transfer function for this self-interacting background has not been calculated, so the coexistence of the onset condition with BBN and structure-formation constraints remains unestablished.

The quartic interaction provides the minimal renormalizable self-interaction considered in the present phenomenological model. 
The ultralight scalar mass also raises a technical-naturalness issue. 
Although $\lambda$ is extremely small in the parameter region considered here, radiative corrections can exceed the tree-level mass if the effective theory is extrapolated to a sufficiently high cutoff scale. We therefore assume that the ultralight mass and the suppression of higher-dimensional operators are protected by an appropriate ultraviolet structure. Constructing an explicit ultraviolet completion is beyond the scope of the present work.

Oscillating ultralight fields can also be probed with atomic clocks~\cite{Arvanitaki:2014faa} or 
pulsar-timing observations
~\cite{Chowdhury:2023xvy}, depending on their couplings. Direct clock signals require additional matter couplings that are not specified in the action considered here.

ULDM can suppress small-scale power and may affect $S_8$~\cite{Rogers:2023ezo,Ye:2021iwa}, whereas some early-time solutions increase it~\cite{Pedrotti:2024kpn}.

The inverse sound-horizon ratio $R_s$ is sensitive to $\tilde m$, with smaller values producing a larger effect. Our calculation illustrates this background mechanism within a sudden-transition approximation; it does not determine an observationally inferred $H_0$ or establish an allowed $\tilde m$ region.
In fact, Eq. (\ref{H0}) is an implicit equation for $H_0$, as $\rho(z)$ on the right-hand side is a function of $H_0$. 
Moreover, our fiducial densities are guided by Planck 2015 base-$\Lambda$CDM results, while $H_{ls}$ and the reference $H_0$ are calculated within our simplified background model.
 Therefore, to reach a definitive conclusion, a full
 likelihood analysis of these cosmological parameters within the self-interacting ULDM model is required, which is beyond the scope of this work.
Recent observations~\cite{Dong:2023jtk,DESI:2025zgx} suggest that dark energy may not be a cosmological constant, which could further complicate future analyses.


\end{document}